\documentclass[aps,prl,twocolumn,showpacs]{revtex4}
\usepackage{dcolumn}
\usepackage{bm}
\usepackage{graphicx}
\usepackage{amsmath}
\usepackage{latexsym}
\usepackage{amsfonts}
\usepackage{amssymb}
\usepackage{array}
\usepackage{epsfig}

\newcommand{\ket}[1]{\left\vert#1\right\rangle}

\newcommand{\sand}[3]{\left\langle#1\vert#2\vert#3\right\rangle}

\begin{document}
\title{Experimental Realization of Deutsch's Algorithm in a One-way Quantum Computer}

\author{M. S. Tame$^1$, R. Prevedel$^2$, M. Paternostro$^1$}
\author{P. B\"ohi$^{2}$}
\altaffiliation[Present address:]{  Max-Planck-Institut f\"ur Quantenoptik und Sektion Physik der Ludwig-Maximilians-Universit\"at, Schellingstr. 4, 80799 M\"unchen, Germany.}
\author{M. S. Kim$^1$}
\author{A. Zeilinger$^{2,3}$}
\affiliation{$^1$School of Mathematics and Physics, Queen's University,~Belfast BT7 1NN, United Kingdom\\
$^2$Faculty of Physics, University of Vienna, Boltzmanngasse 5, A-1090 Vienna, Austria\\
$^3$Institute for Quantum Optics and Quantum Information (IQOQI), Austrian Academy of Sciences, Boltzmanngasse 3, A-1090 Vienna, Austria}

\date{\today}
   
\begin{abstract} 
We report the first experimental demonstration of an all-optical one-way implementation of Deutsch's quantum algorithm on a four-qubit cluster state. All the possible configurations of a balanced or constant function acting on a two-qubit register are realized within the measurement-based model for quantum computation. The experimental results are in excellent agreement with the theoretical model, therefore demonstrating the successful performance of the algorithm.
\end{abstract}

\pacs{03.67.-a, 03.67.Mn, 42.50.Dv, 03.67.Lx}
\maketitle

The increasing interest in topics of quantum information processing (QIP) and quantum computation has stimulated considerable efforts in the realization of {\it quantum hardware} based on various experimental settings. These efforts have resulted in the realization of one and two-qubit logical gates~\cite{nielsenchuang}, even though the networking of these basic building blocks is still far from being practical. Nevertheless, investigations in this direction, both at the experimental and theoretical level are vital for the advancement of QIP. The ultimate aim is the realization of multi-qubit quantum algorithms able to outperform their classical analogues~\cite{nielsenchuang,algoritmi}. In this context, the implementation of few-qubit quantum algorithms represents a step forward in the construction of working processors based on quantum technology~\cite{algorithms,deutsch}. 

Very recently, a radical change of perspective in the design of quantum computational protocols has been proposed and formalized in the ``one-way'' model~\cite{oneway}. Here, computation is not performed by inducing a sequence of logical gates involving the elements of a quantum register, as in the quantum circuit model~\cite{nielsenchuang}. In the one-way case, a multipartite entangled state, the {\it cluster state}, is used as a resource for running a ``program'' represented by single-qubit measurements, performed in order to simulate a given computational task~\cite{oneway}. This new paradigm for quantum computation, which limits the amount of control one needs over a register to the ability of performing single-qubit measurements, has raised an enormous interest in the physical community. It has triggered investigations directed toward a better understanding of the model~\cite{nielsen} and also its practical applications~\cite{nature,varie}. The efforts produced so far have culminated in the experimental demonstration of the basic features of the model, the realization of a two-qubit quantum search algorithm~\cite{nature} and the theoretical proposal for a measurement-based realization of a quantum game~\cite{gioco}. The one-way model is also helping us to understand the paramount role of measurements in the quantum dynamics of a system.

Here, we report the first experimental demonstration of a one-way based implementation of Deutsch's algorithm~\cite{deutsch}. It represents a simple but yet interesting instance of the role that the inherent parallelism of quantum computation plays in the speed-up characterizing quantum versions of classical problems. We have used an all-optical setup, where the construction of cluster states has been successfully demonstrated~\cite{nature,varie}. Negligible decoherence affecting qubits embodied by photonic degrees of freedom ensure the performance of the protocol in a virtually noise-free setting. Although Deutsch's algorithm has been implemented in a linear optical setup before~\cite{resch}, our protocol represents its first realization in the context of one-way quantum computation. It is based on the use of an entangled resource locally equivalent to the cluster state used previously for performing a two-qubit search algorithm~\cite{nature} and reinforces the idea of the high flexibility of cluster resources. We show that four qubits in a linear cluster configuration are sufficient to realize all the possible functions acting on a logical two-qubit register. Two of these result from the application of an entangling gate to the elements of the register. In principle, this gate can be realized by inducing an interaction between the photonic qubits. In our cluster state approach, the required entangling operations are realized by using the entanglement in the cluster resource and the nonlinearity induced by detection. There is no need for {\it engineering} it in a case by case basis~\cite{resch}, which is a very important advantage. The density matrix of the logical output qubits for the functions show excellent performance of the algorithm in our setup.

{\it Model}.- The generalized version of Deutsch's algorithm, also known as the Deutsch-Josza algorithm~\cite{DJ}, takes an $N$-bit binary input $x$ and allows one to distinguish two different types of function $f(x)$ implemented by an oracle. A function is {\it constant} if it returns the same value (either 0 or 1) for all possible inputs of $x$ and {\it balanced} if it returns 0 for half of the inputs and 1 for the other half. Classically one needs to query this oracle as many as $2^{N-1}+1$ times in some cases.
However the quantum version requires only one query in all cases~\cite{DJ}. In the two-qubit version~\cite{deutsch}, the algorithm implements the oracle as a function $f$ on a single query bit $x$ using an input ancilla bit $y$. The applied unitary operation is given by $\ket{x}\ket{y}\to \ket{x}\ket{y\oplus f(x)}$. Preparing the input state as $\ket{+}\ket{-}$, where $\ket{\pm}=(\ket{0}\pm \ket{1})/\sqrt{2}$ and $\{ \ket{0},\ket{1}\}$ is the single-qubit computational basis, the oracle maps the state to $(1/\sqrt{2})[(-1)^{f(0)}\ket{0}+(-1)^{f(1)}\ket{1}]\ket{-}$. By measuring the query qubit in the $\{\ket{\pm}\}$ basis, one can determine which type of function $f(x)$ corresponds to. If $f(x)$ is balanced (constant), the query qubit is always $\ket{-}$ ($\ket{+}$). Thus, only one query of the oracle is necessary, compared to two in the classical version. 
\begin{figure}[t]
\centerline{
\psfig{figure=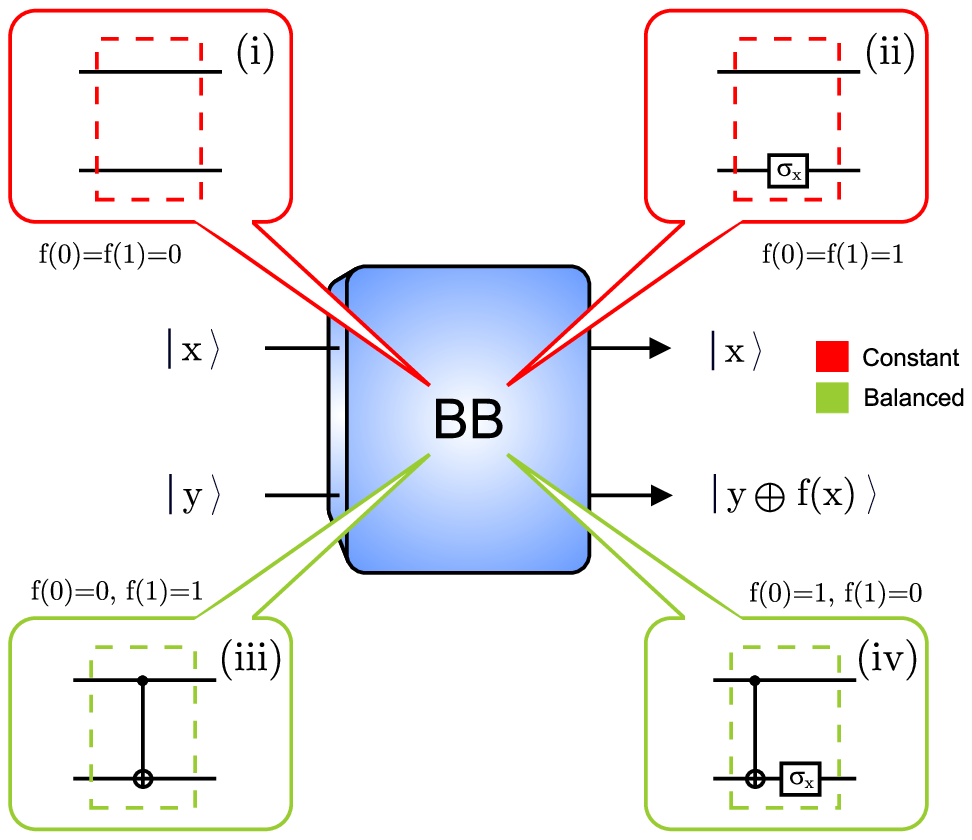,width=4.8cm}}
\caption{Network diagrams for the black boxes in Deutsch's algorithm. We have BB(i)=$\openone \otimes \openone$, BB(ii)=$\openone \otimes \sigma_x$, BB(iii)=${\sf CNOT}$ and BB(iv)=$(\openone \otimes \sigma_x){\sf CNOT}$~(${\sf CNOT}$ denotes a Control-NOT gate).}
\label{fig1}
\end{figure}

The action of the above oracle is either preset or dictated by the outcome of another algorithm. In order to implement all possible configurations that it might take in the two-qubit version, we must be able to construct them using a combination of quantum gates. 
In Fig.~\ref{fig1} we show all possible oracles in terms of their quantum network. By describing each as a ``black box'', one can see that all four black boxes (BB(i)-(iv)) implement their respective oracle operation. In order to carry out the algorithm using these quantum gates, we use a cluster state and carry out one-way quantum computation on it by performing a correct program of measurements. No adjustment to the experimental set-up is necessary.

Given a cluster state, there are two types of single-qubit measurements that allow a one-way quantum computer to operate. First, by measuring a qubit $j$ in the computational basis it can be disentangled and removed from the cluster, leaving a smaller cluster state of the remaining qubits. Second, in order to perform QIP, qubits must be measured in the basis $B_j(\alpha)=\{ \ket{\alpha_+}_j,\ket{\alpha_-}_j \}$, where $\ket{\alpha_{\pm}}_j=(\ket{0}\pm e^{i \alpha}\ket{1})_j/\sqrt{2}$ ($\alpha\!\in\!{\mathbb R}$). Choosing the measurement basis determines the rotation $R_z(\alpha)={\rm exp}(-i \alpha \sigma_z/2)$, followed by a Hadamard operation $H=(\sigma_x + \sigma_z)/\sqrt{2}$ being simulated on an encoded logical qubit in the cluster residing on qubit $j$ ($\sigma_{x,y,z}$ are the Pauli matrices). With a large enough cluster, any quantum logic operation can be performed with a proper choice for the $B_j(\alpha)$'s~\cite{clusterback}.
 
\begin{figure}[b]
\centerline{
\psfig{figure=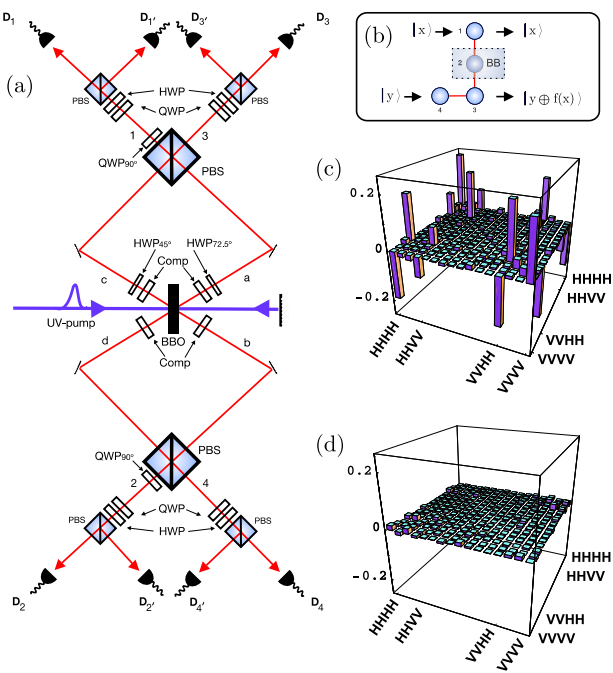,width=8.1cm}}
\caption{{\bf(a)}: Experimental setup. An ultraviolet pump-laser performs two passages through a nonlinear Beta-Barium-Borate crystal (BBO) aligned to produce entangled photon pairs of the form $(\ket{00}-\ket{11})_{ab}/\sqrt{2}$ and $(\ket{00}+\ket{11})_{cd}/\sqrt{2}$. Compensators (Comp) are half-wave plates (HWP) and BBO crystals used in order to counteract walk-off effects at the BBO. By considering the possibility of obtaining a double-pair emission into the same pair of modes and the action of the polarizing-beam splitters (PBS's), the four terms entering $\ket{\Phi_C}$ are obtained and their amplitudes and respective signs adjusted~\cite{nature} with an additional HWP in mode a. The algorithm is executed by using quarter-wave plates (QWPs), HWP's, PBS's and photocounter pairs $\{{\rm D}_j,{\rm D}_{j'} \}$ for the performance of polarization measurements in arbitrary bases of the photons in mode $j$. {\bf (b)}: Sketch of the cluster-state configuration. Qubit $1$ embodies the logical input for $\ket{x}$ and its output. Qubit $4$ (3) is the logical input (output) for $\ket{y}$, which is always found to be $\ket{-}_3$. {\bf (c)} \& {\bf (d)}: Real and Imaginary plots respectively of the reconstructed experimental density matrix $\varrho$.}
\label{fig2}
\end{figure}

{\it Experimental implementation}.- For the entangled resource, in an ideal case, the following four-photon state is produced by means of the set-up shown in Fig.~\ref{fig2}~{\bf (a)} $\ket{\Phi_{c}}=(1/2)(\ket{0000}+\ket{0011}+\ket{1100}-\ket{1111})_{1234}$ with $\ket{0}_{j}$ ($\ket{1}_{j}$) embodied by the horizontal (vertical) polarization state of one photon populating a spatial mode $j=1,..,4$. The preparation of the resource relies on postselection: a four-photon coincidence event at the detectors facing each spatial mode witnesses the preparation of the state. This state is locally equivalent to a four-qubit linear cluster state $\ket{\Phi_{lin}}$ (the local operation being $H_1\otimes\openone_2\otimes\openone_3 \otimes H_4$). The experimentally produced state $\varrho$ is verified by means of a maximum-likelihood technique for tomographic reconstruction~\cite{maxlike} performed over a set of $1296$ local measurements~\cite{nature}, each acquired within a time-window of $500$ s. This provides information about the overall quality of the experimental state on which the algorithm is performed. We have used all the possible combinations of the elements of the mutually unbiased basis $\{\ket{0},\ket{1},\ket{+},\ket{-},\ket{R},\ket{L}\}_{j}$ with $\ket{\pm}_{j}$ embodied by
the polarization state at $\pm{45}^{\circ}$ and $\ket{L/R}_j=(\ket{0}\pm{i}\ket{1})_{j}/\sqrt 2$ corresponding to left and right-circularly polarized photons. This over-complete state tomography has the advantage of providing a more precise state estimation and significantly smaller error bars~\cite{nature}. 
The reconstructed density matrix of $\varrho$ is shown in Fig.~\ref{fig2}~{\bf (c)} \& {\bf (d)} and has a fidelity with the ideal state $\ket{\Phi_C}$ of $F=\sand{\Phi_{c}}{\varrho}{\Phi_{c}}=0.62\pm{0.01}$. The error bar was estimated by performing a 100 run Monte Carlo simulation of the whole state tomography analysis, with Poissonian noise added to the count statistics in each run~\cite{maxlike}. Obtaining a higher fidelity is limited by phase instability during the lengthy process of state tomography and non-ideal optical elements. However, it is well-above the limit $F=0.5$ for any biseparable four-qubit state~\cite{Toth} and demonstrates the presence of genuine four particle entanglement.

In order to perform Deutsch's algorithm on the cluster resource $\ket{\Phi_C}$, we have used a specific set of measurement bases for the qubits in each black box case. In Table~\ref{tab1} we provide these basis sets (BB$_c$) and feed-forward (FF) operations used to carry out the black boxes on $\ket{\Phi_c}$ and also $\ket{\Phi_{lin}}$ (BB basis sets). As BB(ii) and BB(iv) are obtained from BB(i) and BB(iii) by using alternative FF operations (corresponding to adaptive measurements on the output qubits~\cite{clusterback}), in what follows we explicitly describe BB(i) and BB(iii). Fig.~\ref{fig2} {\bf (b)} shows the in-out logical states of the algorithm, where the logical input state corresponding to $\ket{x}=\ket{+}$ is encoded on qubit 1. The state $\ket{y}=\ket{-}$ will be encoded on qubit 3 by measuring qubit 4 in the $B_4(\pi)$ basis during the implementation of the algorithm (described next). This gives $\ket{x}\ket{y}\equiv (\openone \otimes R_z(\pi))\ket{+}\ket{+}$. 

Qubit 2 in $\ket{\Phi_{lin}}$ plays the pivotal role of the oracle as it performs a two-qubit gate on the logical input states $\ket{x}$ and $\ket{y}$. For BB(i), measuring qubit 2 in the computational basis disentangles it from the cluster and $\ket{\Phi_{lin}}$ is transformed into $\ket{\pm}_1(1/\sqrt{2})(\ket{0}\ket{+}\pm\ket{1}\ket{-})_{34}$ ($+$ ($-$) for outcome $\ket{0}_2$ ($\ket{1}_2$)). 
The effective operation performed by this choice of the oracle's measurement basis is $\openone \otimes \openone$. By including the $H$ operation applied to the input state $\ket{y}$ from the measurement of qubit 4, the overall computation results in $(\openone \otimes \openone)(\openone \otimes H R_z(\pi))\ket{+}\ket{+}$ which is equivalent to $\ket{x}\ket{y \oplus f(x)}=(\openone \otimes \openone)\ket{+}\ket{-}$ up to a local rotation $H$ on physical qubit 3, applied at the FF stage. Qubits 1 and 3 can now be taken as the output $\ket{x}\ket{y \oplus f(x)}$.
For BB(iii), upon measuring qubit 2 in the $B_2(\pi/2)$ basis, the oracle applies the gate $(R_z(\pi/2) \otimes R_z(\pi/2)){\sf CPHASE}$ on $\ket{x}$ and $\ket{y}$ (see Tame {\it et al.} in~\cite{nielsen}), where ${\sf CPHASE}$ shifts the relative phase of the state $\ket{1}\ket{1}$ by $\pi$. This gives the computation $\ket{x}\ket{y \oplus f(x)}={\sf CNOT}\ket{+}\ket{-}\equiv (R_z(\pi/2) \otimes R_z(\pi/2)){\sf CPHASE}(\openone \otimes H R_z(\pi))\ket{+}\ket{+}$ up to local rotations $R_z(-\pi/2)\otimes H\,R_z(-\pi/2)$ on qubits 1 and 3, applied at the FF stage. The measurements and outcomes of qubits $1$, $3$ and $4$ constitute the algorithm. The additions to the FF stages described above, together with the measurement of qubit 2 should be viewed as being carried out entirely by the oracle. 
\begin{table}[t]
\begin{ruledtabular}
\begin{tabular}{|c|c|}\hline
 & {\sf Measurement basis} \\ \hline\hline
BB(i) & $\{ B_1(0),\{\ket{0}_2,\ket{1}_2 \},\{\ket{0}_3,\ket{1}_3 \}, B_4(\pi)\}$ \\ \hline
${\rm BB}_{c}$(i) & $\{ \{\ket{0}_1,\ket{1}_1 \},\{\ket{0}_2,\ket{1}_2 \},\{\ket{0}_3,\ket{1}_3 \},\{\ket{1}_4,\ket{0}_4 \} \}$ \\ \hline
BB(iii) & $\{ B_1(\pi/2),B_2(\pi/2),\{\ket{0}_3,\ket{1}_3 \}, B_4(\pi) \}$  \Large{\phantom{A}}\\ \hline
${\rm BB}_{c}$(iii) & $\{ B_1(3 \pi/2),B_2(\pi/2),\{\ket{0}_3,\ket{1}_3 \}, \{\ket{1}_4,\ket{0}_4 \} \}$ \\ \hline
\end{tabular}
\end{ruledtabular}
\caption{Measurement bases for the black boxes. The FF operations are $(\sigma_x^{s_2})_{1} (\sigma_x^{s_4})_{3}$ for ${\rm BB}_{c}$(i) and $(\sigma_z^{s_2\oplus s_4})_{1} (\sigma_x^{s_4})_{3}$ for ${\rm BB}_{c}$(iii). Here, $s_j$ is 0 (1) if the outcome is $\ket{\alpha_+}_j$ ($\ket{\alpha_-}_j$) on qubit $j$.}
\label{tab1}
\end{table}

\begin{figure}[t]
\centerline{
\psfig{figure=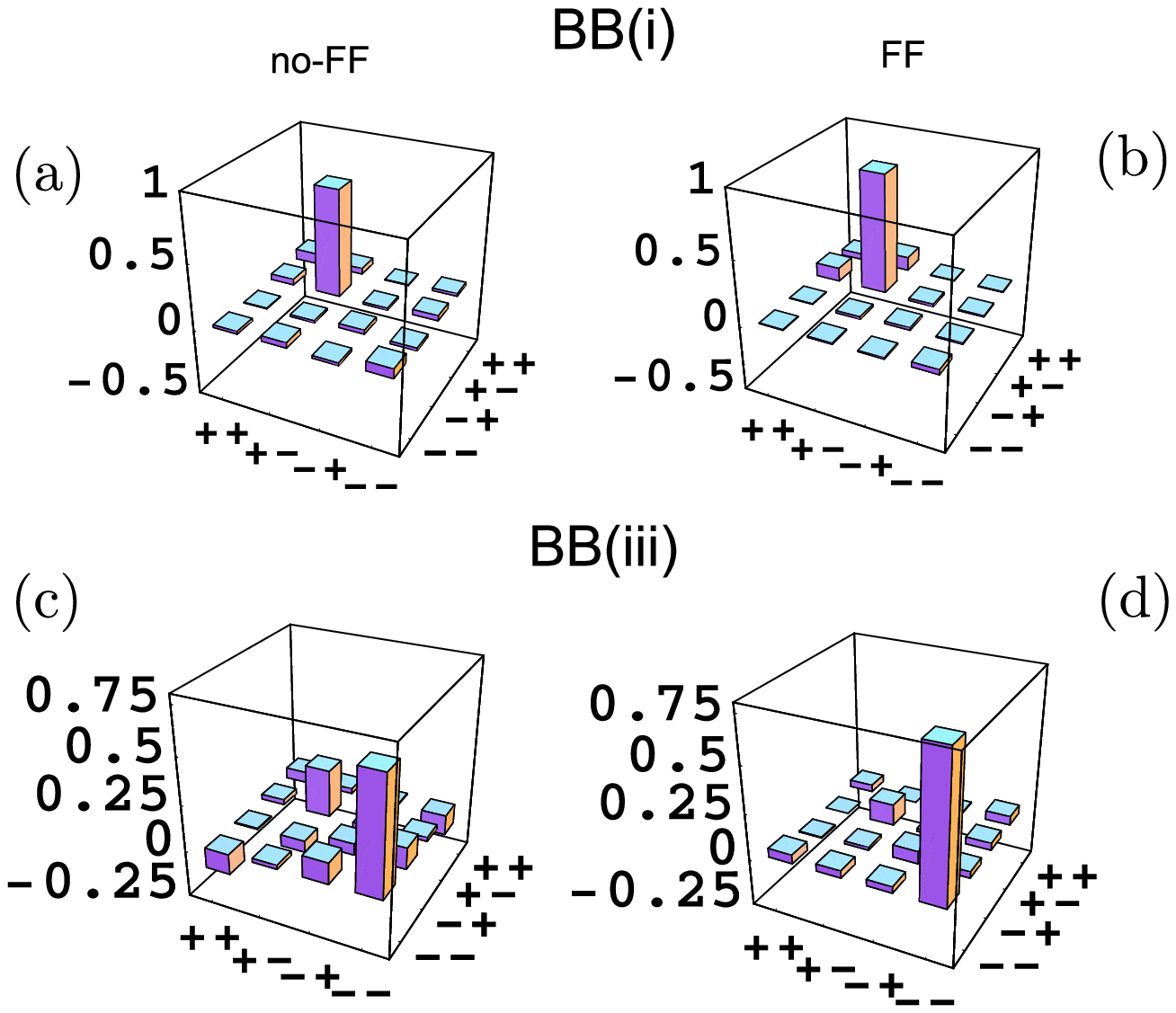,width=6.9cm}}
\caption{The output density matrices for cluster qubits 1 and 3 when BB(i) and BB(iii) are implemented. Panels ${\bf (a)}$ and ${\bf (c)}$ show the real parts of the two-qubit density matrix elements as obtained from a maximum likelihood reconstruction for the no-FF cases of BB(i) and BB(iii) respectively. Panels ${\bf (b)}$ and ${\bf (d)}$ show the corresponding plots for the FF case, due to the randomness of measurement outcomes for qubits $2$ and $4$. In all four cases the imaginary parts 
are zero in theory and negligible in the experiment (average values $< 0.02$).
}
\label{fig3}
\end{figure}

The results of our experiment are shown in Fig.~\ref{fig3}, where we fully characterize the output states of our quantum computer by repeating the algorithm a large number of times. A single run of the algorithm (measuring the output qubit 1 in a specific basis only once) is sufficient in our setup to carry out the quantum computation with success rates as large as $90\%$ ($78\%$) for BB(i) (BB(iii)). However, repeating it several times allows us to verify the density matrix for the quantum state of qubits $1$ and $3$ reconstructed through a maximum likelihood technique~\cite{maxlike}. Although only the logical state residing on qubit 1 provides the outcome of the algorithm, it is useful for the characterization of the quantum computer's performance to also determine the state residing on qubit 3. Ideally, the joint state of qubits 1 and 3 should be the product state $\ket{x}\ket{y \oplus f(x)}$. By obtaining both correct logical output states, we can confirm that the algorithm will run correctly if included in a larger protocol. Fig.~\ref{fig3} shows the output density matrices for BB(i) and BB(iii). Both the no-feed-forward (no-FF) and FF situations are shown. In the latter case, the state of the output qubits is corrected from the randomness of the measurements performed on the physical qubits $2$ and $4$. From the previous analysis, we know that the expected outcome from a single run, when a constant (balanced) function is applied is $\ket{+,-}_{13}$ ($\ket{-,-}_{13}$). Evidently, the reconstructed density matrices, both in the FF and no-FF cases, show a very good performance of the algorithm when compared with the theoretical expectations. The real parts are dominated by the correct matrix elements and no significant imaginary parts are found. Quantitatively, the fidelity with the desired state in the case of a constant (balanced) function is found to be as large as $0.90\pm{0.01}$ ($0.78\pm{0.01}$) for the FF case and $0.82\pm{0.01}$ ($0.63\pm{0.01}$) for the no-FF one.
Moreover, no entanglement is found in any of the joint output states, as witnessed by the negativity of partial transposition criterion~\cite{npt}. The small admixture of the undesired $\ket{+,-}_{13}$ to the expected $\ket{-,-}_{13}$ state when a balanced function is applied (Fig.~\ref{fig3} {\bf (c)}) is due to the non-ideal fidelity of the experimental cluster state with $\ket{\Phi_C}$. This is more pronounced for BB(iii) than for BB(i), where the measurement basis of qubit $2$ breaks the channel between $\ket{x}$ and $\ket{y}$ resulting in a protocol-dependent {\it noise-inheritance} effect for imperfect cluster states (see Tame {\it et al.} in~\cite{nielsen}).
\newline
\phantom{xx}{\it Remarks}.- We have designed, demonstrated and characterized the performance of the first experimental realization of Deutsch's algorithm on a four-qubit cluster state. Our experiment is one of the few quantum algorithms entirely implemented utilizing the one-way model~\cite{nature,gioco}. The agreement between the experimental data and theory is excellent and only limited by the overall quality of the entangled resource in the experiment.
\newline
\phantom{xx}{\it Acknowledgements}.- We thank \v{C}. Brukner, C. Di Franco and A. Stefanov. 
We acknowledge support from DEL, the Leverhulme Trust (ECF/40157), UK EPSRC, FWF, the European Commission under the Integrated Project Qubit Applications (QAP) funded by the IST directorate and the U.S. Army Research Funded DTO Office.


\begin{thebibliography}{99}
\bibitem{clusterback} A detailed introduction to experimental one-way quantum computing can be found in~\cite{nature}.
\end{thebibliography}
\end{document}